\documentclass[aps,prl,twocolumn]{revtex4-1}

\usepackage[final]{pdfpages}

\usepackage{graphicx}  % needed for figures
\usepackage{amssymb}
\usepackage{amsfonts}
\usepackage{amsmath}
\usepackage{bbm}

\newcommand*{\comment}[1]{}

\newcommand{\MHz}{\,\mbox{MHz}}

\newcommand{\hgate}{{\sf H}}
\newcommand{\pgate}{{\sf P}}
\newcommand{\rgate}{{\sf R}}
\newcommand{\xgate}{{\sf X}}

\newcommand{\zgate}{{\sf Z}}
\newcommand{\cnot}{{\sf CNOT}}

\input{qcircuit}

\begin{document}

\title{Quantum computing on encrypted data}

\author{K.A.G. Fisher$^{1,2\star}$}
\email[]{k8fisher@uwaterloo.ca}

\author{A. Broadbent$^{1,3 \star}$}
\email[]{albroadb@iqc.ca}

\author{L.K. Shalm$^{1,4}$}
\email[]{kshalm@uwaterloo.ca}

\author{Z. Yan$^{1,5}$}
\author{J. Lavoie$^{1,2}$}
\author{R. Prevedel$^{1,6}$}
\author{T. Jennewein$^{1,2}$}
\author{K.J. Resch$^{1,2}$}

\affiliation{$\star$These authors contributed equally to this work.}
\affiliation{$^1$Institute for Quantum Computing,  University of Waterloo}
\affiliation{$^2$Department of Physics and Astronomy,  University of Waterloo}
\affiliation{$^3$Department of Combinatorics and Optimization, University of Waterloo, 200 University Avenue West, Waterloo, Ontario N2L 3G1, Canada}
\affiliation{$^4$National Institute of Standards and Technology, Boulder, CO, 80305, USA}
\affiliation{$^5$Centre for Ultrahigh Bandwidth Devices for Optical Systems (CUDOS) \& MQ Photonics Research Centre, Department of Physics \& Astronomy, Macquarie University, Sydney, NSW 2109, Australia}
\affiliation{$^6$Research Institute for Molecular Pathology and Max F. Perutz Laboratories GmbH, Dr.-Bohr-Gasse 7-9, 1030 Vienna, Austria}

\begin{abstract}
The ability to perform computations on encrypted data is a powerful tool for protecting privacy. Recently, protocols to achieve this on classical computing systems have been found. Here we present an efficient solution to the quantum analogue of this problem that enables arbitrary quantum computations to be carried out on encrypted quantum data. We prove that an untrusted server can implement a universal set of quantum gates on encrypted quantum bits (qubits) without learning any information about the inputs, while the client, knowing the decryption key, can easily decrypt the results of the computation. We experimentally demonstrate, using single photons and linear optics, the encryption and decryption scheme on a set of gates sufficient for arbitrary quantum computations. Because our protocol requires few extra resources compared to other schemes it can be easily incorporated into the design of future quantum servers. These results will play a key role in enabling the development of secure distributed quantum systems.
\end{abstract}

\date{\today}
\maketitle

%\section{Introduction}
While quantum computers promise to solve certain classes of problems that are intractable for classical computers\cite{Feynman1982, Deutsch1992, Grover1996, Shor1997}, their development is still in its infancy. It is probable that the first quantum computers will act as servers that potential clients can access remotely. In such a server model, the ability to efficiently implement quantum algorithms on encrypted quantum information is crucial. 
In 2009, the first classical method for fully homomorphic encryption (i.e.~for performing arbitrary computations over encrypted data) was developed\cite{G09}. This enables a client with comparatively little computational power to use an untrusted classical server for performing a computation, without compromising the security of their data. 
Here we have developed the first scheme for carrying out arbitrary computations on encrypted qubits where the client only needs to be able to prepare and send single qubits chosen among a set of four possibilities, and to perform some limited classical communication and computation.
An important feature of our protocol is that during the computation no quantum communication between the client and the server is required. Strictly speaking, fully homomorphic encryption requires that the client's total number of operations be proportional to the size of the input and output only. Our scheme satisfies this requirement at the quantum level, but not at the classical one, since the client's total number of classical operations is proportional to the size of the circuit.  
Nevertheless, our scheme is efficient, requiring only a constant overhead for performing gates on encrypted data, whereas the best known fully homomorphic classical solution\cite{Gentry2012} requires a polylog overhead.

Our protocol (see  Fig.~1) starts with a client who has quantum information that needs to be sent to a remote server for processing. The client first encrypts the input qubits. In the circuit model of quantum computing, a universal set of gates, composed of unitary operations from the Clifford group and one additional non-Clifford gate, is required.  For each non-Clifford gate to be performed in the algorithm, the client must also prepare an auxiliary qubit according to a prescription we will specify. The client sends the encrypted quantum information and the auxiliary qubits to the server, and the server then sequentially performs the gates specified by the quantum algorithm. A round of classical communication between the server and client is required every time a non-Clifford gate is implemented (as shown in Fig.~1h), allowing the client to update the decryption key. After the algorithm is completed, the server returns the encrypted qubits to the client who then decrypts them. Once decrypted, the client has the answer to the computation the server performed while the server has no knowledge about the quantum information it has processed. The server, however, can choose to perform a different computation.  However, for many algorithms of interest\cite{Shor1997}, efficient classical verification methods exist, thus enabling the detection of an incorrect output.

\begin{figure}
\centering
\includegraphics[width=0.9\columnwidth]{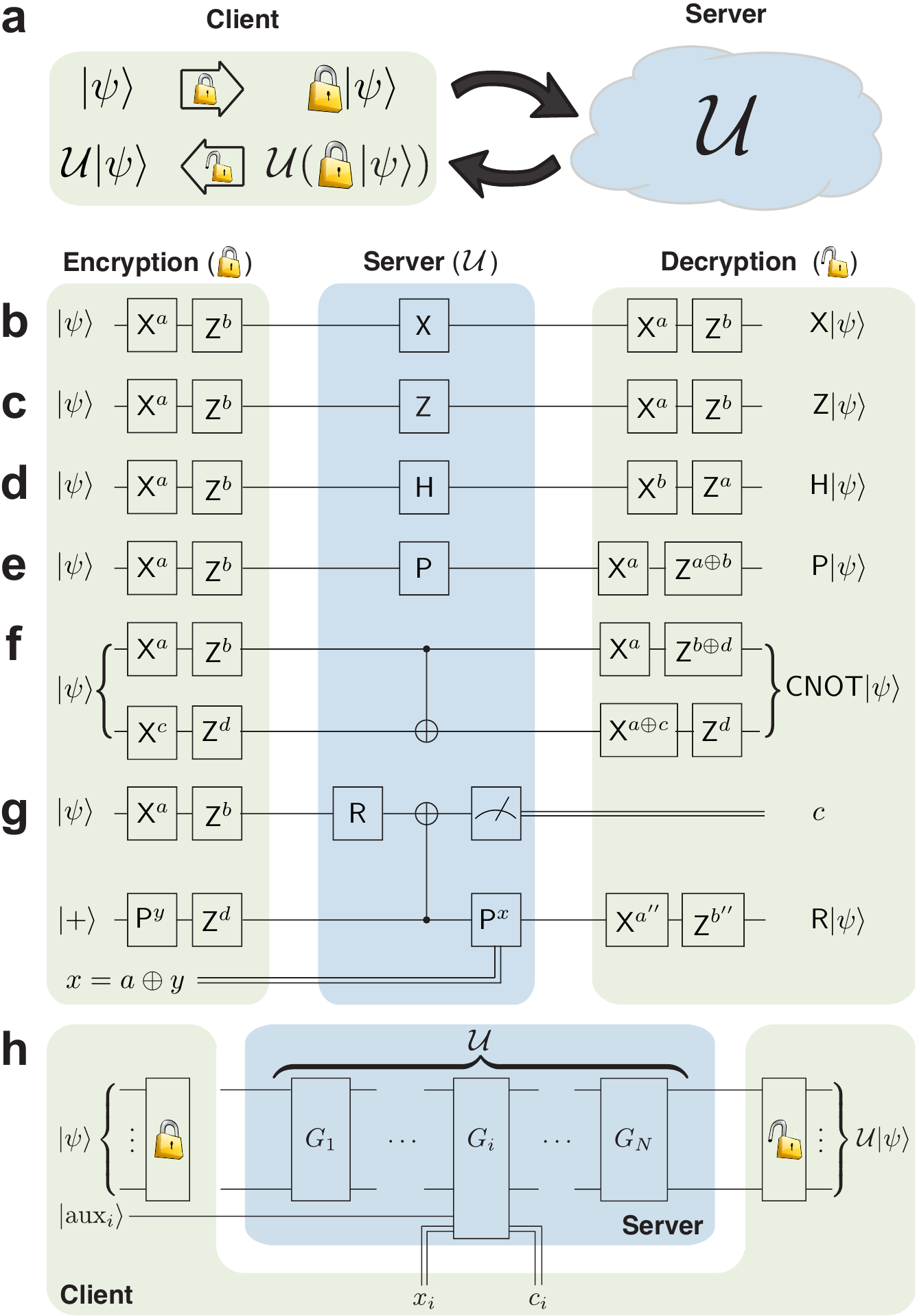}
\caption{ Protocol for quantum computing on encrypted data. 
\textbf{a}, A client encrypts a quantum state $\ket{\psi}$ and sends it to a quantum server, who performs a computation $\mathcal{U}$ on the encrypted qubit. The server returns the state which the client decrypts to get $\mathcal{U} \ket{\psi}$. 
\textbf{b--g}, Encryption and decryption protocols for a universal gate set. Two random classical bits $a, b \in_R \{0,1 \}$ (as well as $c, d \in_R \{0,1 \}$ for the $\cnot$, \textbf{f}) control Pauli rotations $\xgate$  and $\zgate$ to encrypt state $\ket{\psi}$. 
\textbf{b--f}, Clifford gates do not require any additional resources, and decryption is straightforward. 
\textbf{g}, The non-Clifford $\rgate$ gate requires the client to send an auxiliary qubit $\zgate^d \pgate^y \ket{+}$, where $y,d \in_R \{0,1\}$, to control a $\cnot$ gate with the encrypted qubit. 
The server measures the encrypted qubit and outcome $c \in \{0,1\}$ is returned to the client, which is used in decryption. The client sends a single classical bit, $x= a\oplus y$, to control a $\pgate$ gate on the auxiliary qubit, which is returned to the client as $\xgate^{a''} \zgate^{b''} \rgate \ket{\psi}$, where $a'' = a \oplus c$ and $b'' = a(c \oplus y \oplus 1) \oplus b \oplus d \oplus y$.
\textbf{h}, For a computation, the client encrypts and sends $\ket{\psi}$ to be processed, as well as auxiliary qubits, $\ket{\text{aux}_i}$, for any $\rgate$ gates in the computation. The server performs a series of gates $\mathcal{U} = G_N\cdots G_1$. Communication is only needed when gate $G_i$ is an $\rgate$, and then only with classical bits. Processed qubits are returned to the client for decryption.
 }
\label{fig:protocol}
\end{figure}

Our scheme is part of a rapidly developing field that tackles the problem of secure delegated quantum computation. There have been several novel approaches to this problem, including hiding a circuit from the remote quantum server \cite{BFK09, BKBFZW12}, computing on encrypted quantum data using multiple rounds and bits of quantum communication\cite{C05, AharonovBE10, DNS12, BGS12}, and sophisticated methods that provide an additional verification mechanism\cite{AharonovBE10, DNS12, BGS12} (see Supplementary Table 1 for more details). 
While some of these schemes, in principle, can be used to accomplish similar outcomes as our protocol, they can lead to very different client-server relationships in practice. For example, a recent experiment used the measurement-based model of quantum computing to demonstrate the complementary problem of hiding from a server the circuit that is to be performed\cite{BFK09, BKBFZW12}. This method, known as blind quantum computing, can be extended to compute on encrypted data, but would require more than eight times as many auxiliary qubits and significantly more rounds of classical communication. Furthermore, blind computation uses random qubits chosen from a set of eight possibilities---our contribution reduces this to just four. 

More fundamentally, blind quantum computing demands a very different relationship between the client and server as compared to our approach that is inspired by homomorphic encryption. In the blind model, the client must provide both the hidden algorithm to be performed and the encrypted data to be computed on; in our scheme the client provides only the data while the server provides the agreed upon algorithm. Our protocol mirrors the client-server relationships that exist today where a server is free to focus on iterating and improving the algorithms they provide. This frees the client from needing to develop and optimize the algorithms they use, while the server is able to specialize in providing targeted services. In the blind model this division of labour does not exist; the server is treated as a ``dumb'' resource while the client is fully responsible for maintaining and supplying the algorithms. While there are many scenarios where carrying out blind quantum computing is desirable, our protocol enables secure delegated quantum networks to develop that closely resemble today's client-server relationships. 

In our scheme, to encrypt a qubit $\ket{\psi}$, a client applies a combination of Pauli $\xgate$ and $\zgate$ operations: 
\begin{equation}
\label{E:Encryption}
\xgate^a \zgate^b \ket{\psi} = \ket{\psi}_\text{encrypted},
\end{equation}
where $a$ and $b$ are randomly assigned to the values of $0$ or~$1$ and form the key. The action of the encryption maps the initial state of the qubit to one of four possible final states, which sum to the completely mixed state; as long as the values $a$ and $b$ are used only once, this is the quantum equivalent\cite{AMTW00} of the classical one-time pad. Knowing $a$ and $b$, it is possible to decrypt the state by reversing the $\xgate$ and $\zgate$ rotations. The Clifford gates we study include the single-qubit Pauli $\xgate$ and $\zgate$ rotations, the two-qubit controlled-NOT ($\cnot$) gate, and the single-qubit Hadamard, $\hgate \ket{j} \mapsto \frac{1}{\sqrt{2}}( \ket{0} + (-1)^j \ket{1})$, and phase, $\pgate \ket{j} \mapsto \left ( e^{i \pi /2} \right )^j \ket{j}$, gates where $j \in \{ 0,1 \}$\cite{Nielsen2000}. The actions of the Clifford gates on an encrypted qubit are straightforward due to their commutation relations with the Pauli operators (see Fig.~1b--f), and do not require any additional classical or quantum resources\cite{C05}. The client only needs to know what gates are being carried out to update the knowledge of the decryption key. 

\begin{figure}[t!]
\centering
\includegraphics[width=1\columnwidth]{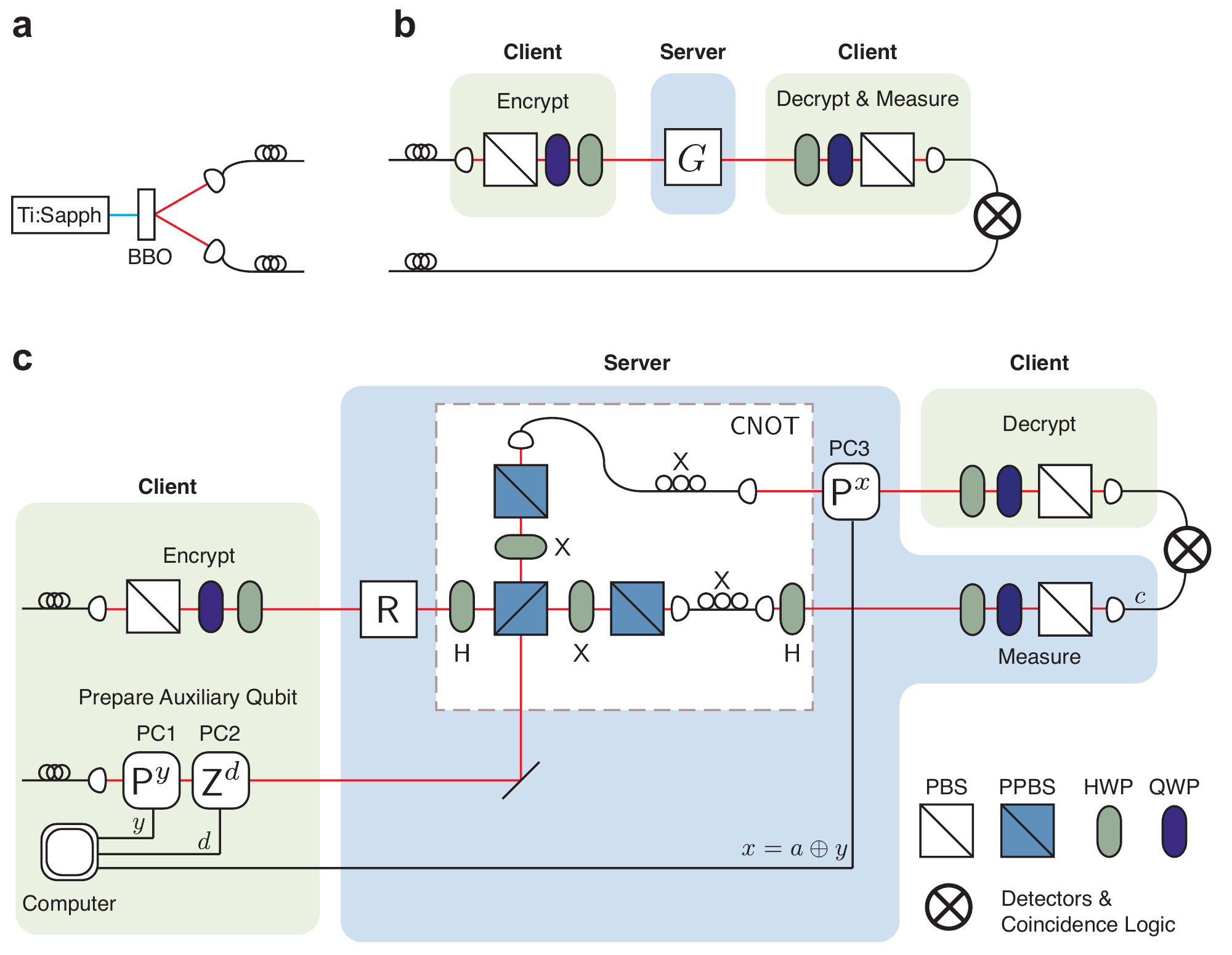}
\caption{ Experimental setup.
 \textbf{a}, Photon pairs are generated via SPDC using a frequency-doubled Ti:Sapph laser to pump a barium borate (BBO) crystal. Photons are coupled into single-mode fibres. 
\textbf{b}, The client prepares and encrypts the qubit $\ket{\psi}$, upper rail, using a PBS, QWP and HWP, and then sends it to the server.
Single-qubit Clifford gates, shown as $G$, are implemented as follows:  $\xgate$ is a HWP at $45^\circ$; $\zgate$ is a HWP at $0^\circ$; $\hgate$ is a HWP at $22.5^\circ$; $\pgate$ is a QWP at $0^\circ$. The photon is returned to the client, where it is measured using a HWP, QWP and PBS, and detected in coincidence with the second photon from the SPDC pair (lower rail).
\textbf{c}, The client prepares and encrypts $\ket{\psi}$, upper rail, as in \textbf{b}. The client also prepares an auxiliary photon, lower rail, to one of $\{ \ket{+}, \ket{-}, \ket{+_y}, \ket{-_y} \}$ using Pockels' cells (PC1, PC2) triggered by randomly generated classical bits $y$ and $d$. The $\rgate$ gate, a tilted HWP at $0^\circ$, acts on photon $\ket{\psi}$. Both photons pass through the $\cnot$, where they interfere at a PPBS. The encrypted photon $\ket{\psi}$, in the lower rail, is measured by the server after the $\cnot$, and the outcome $c$ is used by the client in decryption. The auxiliary photon, now in the upper rail, passes through a third Pockels' cell (PC3), performing $\pgate^x$, where $x = a \oplus y$ is a classical bit sent from the client, and is returned to the client for decryption and measurement. To test the $\cnot$ gate Pockels' cells are not used, and state preparation and measurement apparatuses are used in both arms.
}
\label{fig:setup}
\end{figure}

\begin{figure}[h!]
\centering
\includegraphics[width=\columnwidth]{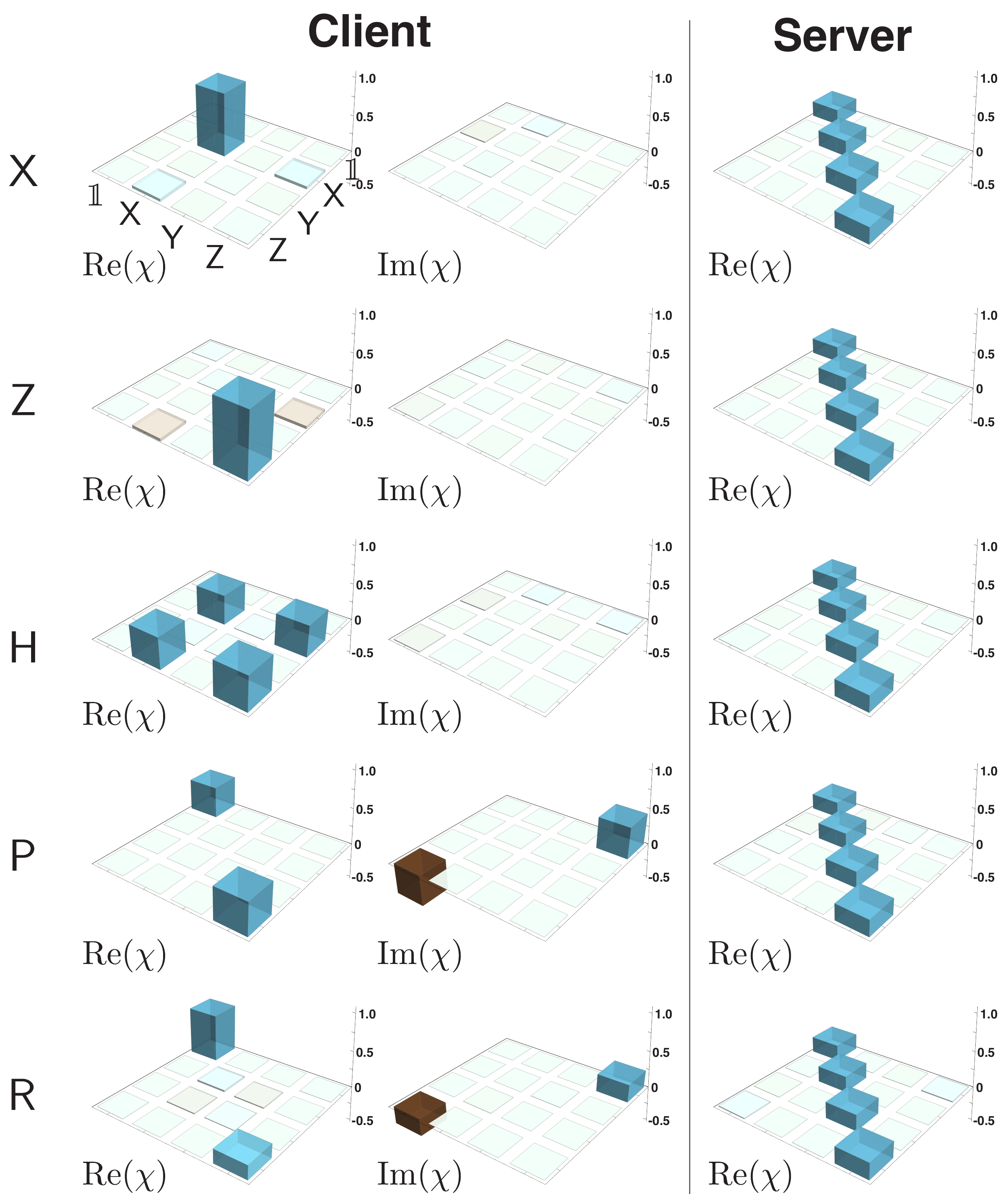}
\caption{ Single-qubit gate results. 
The left panel, the client, shows real and imaginary parts of reconstructed $\chi$ matrices (expressed in the basis of Pauli operators) for the single-qubit gates when decrypted. Fidelities with ideal $\xgate, \zgate, \hgate, \pgate,$ and $\rgate$ gates are $0.984 \pm 0.002$, $0.985 \pm 0.001$, $0.983 \pm 0.001$, $0.985 \pm 0.001$, $0.863 \pm 0.004$, respectively. The right panel, the server, shows the real parts (imaginary parts were negligible) of the reconstructed $\chi$ matrices when not decrypted, all giving process fidelities of $\mathcal{F} \geq 0.999$ with the completely depolarizing channel. Ideal $\chi$ matrices are shown in the Supplementary Information. }
\label{fig:results}
\end{figure}

\begin{figure*}[t]
\centering
\includegraphics[width=1.5\columnwidth]{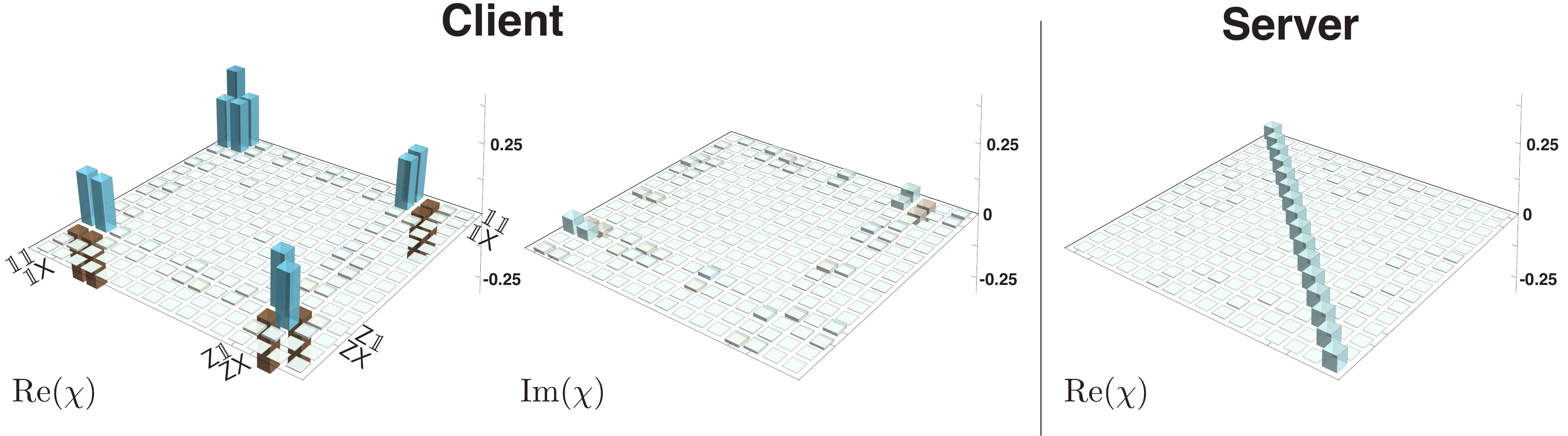}
\caption{ $\cnot$ gate results. 
The left panel, the client, shows real and imaginary parts of the reconstructed $\chi$ matrix for the $\cnot$ gate when the decryption key is known. Fidelity with the ideal $\cnot$ gate is $0.869 \pm 0.004$. The right panel, the server, shows the real part (imaginary part is negligible, $(<0.004)$) of the $\chi$ matrix when the key is unknown. The process fidelity with the completely depolarizing channel is $0.996\pm 0.001$. Ideal $\chi$ matrices are shown in the Supplementary Information. }
\label{fig:CNOTresults}
\end{figure*}

Clifford gates alone are insufficient for universal quantum computing\cite{Gott98}; at least one non-Clifford gate is required. We study the non-Clifford $\rgate$ gate, which has the following action: $\rgate \ket{j} \mapsto \left ( e^{i \pi /4} \right )^j \ket{j} $ for $j \in \{ 0,1 \}$. Performing the $\rgate$ gate on encrypted qubits is not trivial as it does not commute through the encryption in the same simple manner as the Clifford gates. 
This is because the server, when applying the $\rgate$ gate, can introduce an error, equivalent to applying an extra $\pgate$ gate, when $a=1$: $\rgate \xgate^a \zgate^b \ket{\psi}  = \xgate^a \zgate^{a \oplus b} \pgate^a  \rgate \ket {\psi} $.
To prevent the client from needing to divulge the value of $a$, compromising the security of the computation, the server implements a hidden $\pgate$ gate that is controlled by the client (see Fig.~1g).
To do this, before the server begins the computation the client sends as many auxiliary qubits as there are $\rgate$ gates in the circuit. These auxiliary qubits are encoded as $\pgate^y \zgate^d \ket{+}$ with $y,d \in_R \{0,1 \}$, resulting in one of the four following states that lie along the equator of the Bloch sphere: $ \{ \ket{+}= \frac{1}{\sqrt{2}} (\ket{0} + \ket{1}), \ket{-}= \frac{1}{\sqrt{2}} (\ket{0} - \ket{1}), \ket{+_y}= \frac{1}{\sqrt{2}} (\ket{0} + i \ket{1}), \ket{-_y}= \frac{1}{\sqrt{2}} (\ket{0} - i \ket{1})  \}$. These are the four standard BB84 states\cite{BB84} rotated to a different basis. 
After the server implements an $\rgate$ gate, it then performs a $\cnot$ between one of the auxiliary qubits and the encrypted state  $\rgate \xgate^a \zgate^b \ket{\psi}.$ The server measures the encrypted qubit in the computational basis, and returns the outcome $c$ to the client to update the decryption key. After the $\cnot$, the state of the auxiliary qubit is $\xgate^{a'} \zgate^{b'} \pgate^{a \oplus y} \rgate \ket{\psi}$; the extra unwanted phase gate now depends on the values of both $a$ and $y$ which only the client knows.  The client sends a single classical bit, $x = a \oplus y$, which controls whether the server implements an additional corrective $\pgate$ gate without ever revealing the value of $a$. The final state is then $\xgate^{a''} \zgate^{b''}  \rgate \ket{\psi}$ as desired, and the decryption key bits, $a''$ and $b''$, now depend on the values of $a$, $b$, $c$, $d$, and $y$ as shown in Fig.~1g. 
This solution is inspired by circuit manipulation techniques\cite{ZLC00, CLN05}. The Supplementary Information provides a detailed proof. It also provides a novel simulation-based security definition, applicable to any untrusted server sharing arbitrary prior information with the client, and a proof via an entanglement-based protocol\cite{SP00}.

We implement a proof-of-principle of the protocol using linear optics. The state of the qubit is encoded into the polarization of single photons with the horizontal and vertical polarizations representing $\ket{0}$ and $\ket{1}$ respectively. Single photons are generated (see Fig.~2a) via spontaneous parametric downconversion (SPDC). The state preparation and encryption, $\xgate^a \zgate^b \ket{\psi}$, are carried out using a polarizing beamsplitter (PBS), quarter-waveplate (QWP) and half-waveplate (HWP), and the single-qubit Clifford gates are implemented using waveplates (see Fig.~2b). The $\cnot$ gate (see Fig.~2c) is implemented using two-photon interference\cite{HOM1987} at a partially-polarizing beamsplitter (PPBS)\cite{Kiesel2005, Langford2005, Okamoto2005}, which fully transmits horizontally polarized light, but reflects $2/3$ of the vertical polarization.

To implement the $\rgate$ gate on an encrypted qubit we use an auxiliary qubit along with the $\cnot$ as shown in Fig.~2c. The auxiliary qubit is randomly prepared by the client in one of the four rotated BB84 states,  $\pgate^y \zgate^d \ket{+}$, using waveplates and Pockels' cells as fast optical switches\cite{Pittman2002, Prevedel2007, Ma2012} (see Methods), and then sent to the server. The Pockels' cells are switched at 1MHz -- two orders of magnitude faster than our singles rate from SPDC. This means that the probability of more than one photon being present for each Pockels' cell setting is very small, making negligible the amount of information the server can obtain about the state of the auxiliary qubit, and hence the value of~$y$. The server first performs an $\rgate$ gate on the encrypted qubit followed by a  $\cnot$ with the auxiliary qubit.  The client then sends the server a classical bit, $x = a \oplus y$, which controls whether the server implements an additional corrective $\pgate$ gate using a third Pockels' cell. Finally, the server returns to the client the encrypted auxiliary qubit containing the final state for processing.

In order to characterize our gates we use quantum process tomography (QPT)\cite{Nielsen2000, Poyatos1997, OBrien2004, Chow2009}; this provides us with complete information, in the form of a process matrix $\chi$,  about how each gate acts on and transforms an arbitrary input state. The client first prepares a set of encrypted input states that the server acts on, and then the client performs measurements on the outputs. For our single-qubit gates the client prepares an overcomplete set of inputs that are the eigenstates of the Paulis $\{ \ket{0}, \ket{1}, \ket{+}, \ket{-}, \ket{+_y}, \ket{-_y} \}$. Our encryption scheme, $\xgate^a \zgate^b \ket{\psi}$, maps each of these Pauli eigenstates into one another. After the server processes the gate, the client performs measurements in each Pauli basis. By choosing this set of input states, and keeping track of the values of $a$ and~$b$, the client is able to completely characterize the action of the gate over all possible encryptions. Similarly, for the two-qubit $\cnot$ gate the client prepares and measures all 36 eigenstates of the tensor products of the Paulis $\{\ket{00}, \ket{01}, ..., \ket{-_y -_y} \}$. Again, the encryption scheme maps each of the input eigenstates of the Pauli tensor products into one another, allowing all encryption possibilities to be studied.

The client, knowing the decryption key, is able to decrypt and post-process the tomography data. The results for the decrypted single-qubit gates are shown in Fig.~3 and the results for the $\cnot$ are shown in Fig.~4. The fidelities\cite{Nielsen2000} of the $\xgate$, $\zgate$, $\hgate$, $\pgate$, $\rgate$ and $\cnot$ gates are $0.984 \pm 0.002$, $0.985 \pm 0.001$, $0.983 \pm 0.001$, $0.985 \pm 0.001$, $0.863 \pm 0.004$ and $0.869 \pm 0.004$ respectively. Loss of fidelity for single-qubit Clifford gates is predominately due to coherent noise, i.e. over- or under-rotation of a unitary, meaning that multiple gates can be performed in sequence maintaining high fidelity. 
Loss of fidelity for the $\cnot$ and $\rgate$ gates originates from emitted double pairs in the photon source and mode mismatch at the main PPBS.
From the client's perspective, the server has performed the correct computations on the encrypted inputs. However, if the decryption keys are not known, then each gate acts as a completely depolarizing channel that leaves input qubits in the maximally mixed state (as shown in Fig.g~3 and Fig.~4). The process matrices were then reconstructed from the same data as before, but without decryption. Each case had high fidelity with the completely depolarizing channel: $0.999 \pm 0.001 $ for the single-qubit gates, and $0.996 \pm 0.001$ for the $\cnot$. Without knowledge of the decryption keys, the server gains no information about the state. 
For a more detailed analysis of the experimental security of our implementation see the Supplementary Information.

In information security often the weakest link is not the transmission of encrypted data, but, rather, security breaches at the end points where the data is no longer encrypted. A major advance of our scheme is that it eliminates one of the end points as a security risk; a remote server no longer needs to decrypt the quantum information in order to process it and carry out computations. The overhead in quantum resources required to compute on encrypted quantum data is so low (only one auxiliary qubit per non-Clifford gate) that it will be straightforward for future quantum servers to incorporate our protocol in their design, dramatically enhancing the security of client-server quantum computing; our protocol has even less overhead than the best classical fully homomorphic encryption scheme, and provides information-theoretic (as opposed to just computational) security. This method for computing on encrypted quantum data, combined with the techniques developed for quantum circuit hiding\cite{BFK09, BKBFZW12}, form a complete security system that will enable secure distributed quantum computing to take place, ensuring the privacy and security of future quantum networks.

\textbf{Methods.}
In the $\rgate$ gate protocol, we initialize auxiliary photons to one of the four $\{\ket{+}, \ket{-}, \ket{+_y}, \ket{-_y} \}$ states using Pockels' cells. A Pockels' cell performs a fast-switching unitary operation triggered by applying a strong electric field which rapidly changes the index of refraction of a nonlinear medium; here the medium is rubidium titanyl phosphate, RbTiOPO$_4$ (RTP). The values of bits $y$ and $d$ are randomly generated by a computer, and a trigger circuit (based on a self-built CPLD design) is used to drive the Pockels' cells at a rate of $1 \MHz$. Single-photon rates are reduced to $\mathord{\sim}3800$~Hz to limit the probability of two auxiliary photons being present in the Pockels' cells during a single setting of $y$ and $d$. Reduced rates also limit the effect of emitted double pairs on the fidelity of the $\cnot$ operation. Photons are detected using silicon avalanche photo-diodes (PerkinElmer four-channel SPCM-AQ4C modules), and coincidence photon events are recorded using a custom design coincidence logic.
For all gates, the process that the server observed was attained by summing the measured counts over the all the encryption cases $a,b \in \{ 0, 1 \} $. For example, if the client input the state $\ket{0}$, then the server, not knowing the encryption key, would half the time assume $\ket{1}$ was input and sort the measured counts accordingly. 
For the $\rgate$ gate the client decrypts by sorting photon counts into 8 bins based on the values of $y$, $d$ and~$c$. The server, not knowing values of $y$ and~$d$, could at most sort counts into two bins based on $c$, and observes a maximally mixed state due to the active switching, before summing over the encryption key cases.
QPT was performed using a maximum likelihood technique\cite{James2001, Chow2009}. Uncertainties in these values are found by adding Poissonian noise to the measured photon counts and performing 100 Monte Carlo iterations of the $\chi$ matrix reconstructions.

\bibliography{arxiv_refs}

\begin{acknowledgements}
\textbf{Acknowledgements: } 
We are grateful for financial support from Ontario Ministry of Research and Innovation ERA, QuantumWorks, NSERC, OCE, Industry Canada and CFI. A.B., L.S. and T.J. acknowledge the support of the Canadian Institute for Advanced Research. R.P. acknowledges support from the FWF (J2960-N20), MRI, the VIPS Program of the Austrian Federal Ministry of Science and Research and the City of Vienna as well as the European Commission (Marie Curie, FP7-PEOPLE-2011-IIF). A.B. is grateful for Serge Fehr for pointing out the proof technique of ref. [18] and its applicability to our scenario. 
 \textbf{Author Contributions: }
 A.B. designed the protocol and proved its security. K.F., L.S., R.P. and K.R. conceived the experiment. K.F. conducted the experiment with the help of J.L. and Z.Y. and under the supervision of K.R. and T.J. The first draft of the manuscript was written by K.F. and L.S. All authors contributed to the final draft.
% \textbf{Correspondence: } Correspondence and requests for materials should be addressed to K.F.  (email: \\k8fisher@uwaterloo.ca), to A.B. (email:  albroadb@iqc.ca), or to L.S. (email: kshalm@uwaterloo.ca).
\end{acknowledgements}

%%%%%%%%%%%%%%%%%%%%%%%%%%%%%%%%%%%%%%%%

%Defintions for Supplementary
\onecolumngrid
\appendix
\newpage


\section*{Supplementary material}

# Supplementary Information
# Quantum Computing on Encrypted Data

K.A.G. Fisher[†], A. Broadbent[⋆], L.K. Shalm, Z. Yan,
J. Lavoie, R. Prevedel, T. Jennewein, K.J. Resch

Institute for Quantum Computing
University of Waterloo

⋆ albroadb@iqc.ca, † k8fisher@uwaterloo.ca,

# Contents

| Secure Assisted Quantum Computation [Chi05] | This Work |
|---|---|
| $\mathcal{O}(s)$ rounds of quantum communication | One round of quantum communication |
| Clients performs quantum SWAP gate | Client performs no two-qubit gates |
| **Quantum Prover Interactive Proof [ABOE10]** | **This Work** |
| Client needs constant-sized quantum computer | Client's quantum power limited to encryption and preparing random BB84 states |
| Verification of result | No verification of result |
| **Universal Blind Quantum Computing [BFK09]** | **This Work** |
| Each gate (including identity) uses:<br>• 8 auxiliary qubits (chosen out of 8 possibilities)<br>• 32 bits of classical communication | • Clifford group gates are non-interactive<br>• R-gate requires a single auxiliary qubit (chosen out of 4 possibilities) and 1 bit of classical communication in each direction. |

**Table 1** – Comparison with related work

# 1 Universal Circuits

Let $\mathcal{C}$ be a collection of quantum circuits acting on $n$-qubits. A quantum circuit $\mathcal{U}$ on $n + m$ qubits is *universal for* $\mathcal{C}$ if, for every circuit $C \in \mathcal{C}$, there is a string $x \in \{0,1\}^m$ (the *encoding*) such that for all inputs $y \in \{0,1\}^n$, $\mathcal{U}(|y\rangle \otimes |x\rangle) = C(|y\rangle) \otimes |x\rangle$.

Given such a $C \in \mathcal{C}$, we apply our scheme for computing on encrypted data, where the circuit is $\mathcal{U}$ and we append to the input the basis state $|x\rangle$ before encryption. We thus achieve hiding of not only the data, but also the value $x$, which represents the index of $C$ within $\mathcal{C}$.

The complexity (in terms of communication and auxiliary qubits) of the resulting scheme for a fixed $C$ depends on the collection $\mathcal{C}$ that is chosen, together with the choice of $\mathcal{U}$. Clearly, the complexity can only increase compared to executing $C$ on encrypted data. A typical scenario would be to hide a circuit $C$ consisting of $g$ gates within all circuits of the same configuration, but where each gate position consists of H, R, CNOT or the identity[BFK09]. A solution would be for $\mathcal{U}$ to act on $n + 3g$ qubits, with each gate being controlled-versions of H, R and CNOT where the control-bits are part of the encoding, $x$. The straightforward solution based on [NC00] and [BFGH10] requires twenty-four R gates per position. Interesting open questions are improving this scheme, as well as determining efficient universal circuits for fixed collections of quantum circuits (see [SR07, BFGH10]).

# 2 Comparison with other schemes

Previous results have achieved similar functionality, but require more resources. Comparisons with related work are summarized in Table 1, where $s$ in the size of the circuit.

# 3 Correctness of the R-gate protocol

We give below a step-by-step proof of the correctness of the R-gate protocol as given in Figure 1g. The basic building block is the circuit identity for an X-teleportation from [ZDC00], which we re-derive here. Also of relevance to this work are the techniques developed by Childs, Leung, and Nielsen [CLN05] to manipulate circuits that produce an output that is correct *up to known Pauli corrections*.

We will make use of the following identities which all hold up to an irrelevant global phase: XZ = ZX, PZ = ZP, PX = XZP, RZ = ZR, RX = XZPR, $P^2$ = Z and $P^{a \oplus b} = Z^{a \cdot b} P^{a+b}$ (for $a, b \in \{0, 1\}$).

1. Our first circuit identity swaps a qubit $|\psi\rangle$ with the state $|+\rangle$ and is easy to verify.

2. We can measure the top qubit in the above circuit and classically control the output correction. We have thus re-derived the circuit corresponding to the "X-teleportation" of [ZDC00].

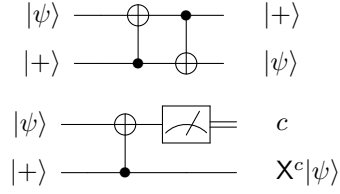

3. Next, we re-define the input to be $\mathsf{R}\mathsf{X}^a\mathsf{Z}^b|\psi\rangle$, so the output becomes $\mathsf{X}^c\mathsf{R}\mathsf{X}^a\mathsf{Z}^b|\psi\rangle = \mathsf{X}^{a\oplus c}\mathsf{Z}^{a\oplus b}\mathsf{P}^a\mathsf{R}|\psi\rangle$.

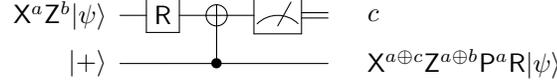

4. Then add three gates ($\mathsf{P}^y$, $\mathsf{Z}^d$, $\mathsf{P}^{a\oplus y}$) to the bottom wire (see circuit below). Using the fact that the $\mathsf{P}$ and $\mathsf{Z}$ commute with control, and applying identities given above, we get as output what we expect:

$$\mathsf{P}^{a\oplus y}\mathsf{Z}^d\mathsf{P}^y\mathsf{X}^{a\oplus c}\mathsf{Z}^{a\oplus b}\mathsf{P}^a\mathsf{R}|\psi\rangle = \mathsf{Z}^{a\cdot y}\mathsf{P}^{a+y}\mathsf{Z}^d\mathsf{P}^y\mathsf{X}^{a\oplus c}\mathsf{Z}^{a\oplus b}\mathsf{P}^a\mathsf{R}|\psi\rangle \tag{1}$$
$$= \mathsf{Z}^{d\oplus a\cdot y\oplus y}\mathsf{P}^a\mathsf{X}^{a\oplus c}\mathsf{Z}^{a\oplus b}\mathsf{P}^a\mathsf{R}|\psi\rangle \tag{2}$$
$$= \mathsf{Z}^{d\oplus a\cdot y\oplus y}\mathsf{X}^{a\oplus c}\mathsf{Z}^{a(a\oplus c)}\mathsf{P}^a\mathsf{Z}^{a\oplus b}\mathsf{P}^a\mathsf{R}|\psi\rangle \tag{3}$$
$$= \mathsf{X}^{a\oplus c}\mathsf{Z}^{d\oplus a\cdot y\oplus y\oplus a^2\oplus a\cdot c}\mathsf{Z}^b\mathsf{R}|\psi\rangle \tag{4}$$
$$= \mathsf{X}^{a\oplus c}\mathsf{Z}^{a(c\oplus y\oplus 1)\oplus b\oplus d\oplus y}\mathsf{R}|\psi\rangle \tag{5}$$

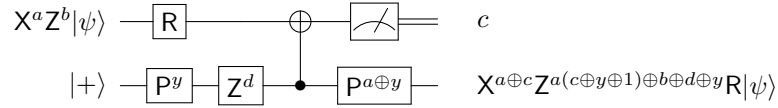

## 4  Security definition and proof

### 4.1  Preliminaries

#### 4.1.1  Quantum registers and channels

A *quantum register* is a collection of qubits in some finite dimensional Hilbert space, say $\mathcal{X}$. We denote $\mathrm{D}(\mathcal{X})$ the set of density operators acting on $\mathcal{X}$. The set of all linear mappings from $\mathcal{X}$ to $\mathcal{Y}$ is denoted by $\mathrm{L}(\mathcal{X}, \mathcal{Y})$, with $\mathrm{L}(\mathcal{X})$ being a shorthand for $\mathrm{L}(\mathcal{X}, \mathcal{X})$. A linear super-operator $\Phi : \mathrm{L}(\mathcal{X}) \to \mathrm{L}(\mathcal{Y})$ is *admissible* if it is completely positive and trace-preserving. Admissible super-operators represent mappings from density operators to density operators, that is, they represent the most general quantum maps.

Given admissible super-operators $\Phi$ and $\Psi$ that agree on input space $\mathrm{L}(\mathcal{X})$ and output space $\mathrm{L}(\mathcal{Y})$, we are interested (for cryptographic purposes) in characterizing how "indistinguishable" these processes are. The *diamond norm* provides such a measure: given that $\Phi$ or $\Psi$ is applied with equal probability, the optimal procedure to determine the identity of the channel with only one use succeeds with probability $1/2 + \|\Phi - \Psi\|_\diamond / 4$. Here,

$$\|\Phi - \Psi\|_\diamond = \max\{\|(\Phi \otimes \mathbb{1}_\mathcal{W})(\rho) - (\Psi \otimes \mathbb{1}_\mathcal{W})(\rho)\|_1 : \rho \in \mathrm{D}(\mathcal{X} \otimes \mathcal{W})\}, \tag{6}$$

where $\mathcal{W}$ is any space with dimension equal to that of $\mathcal{X}$ and $\mathbb{1}_\mathcal{W}$ is the identity in $\mathrm{L}(\mathcal{W})$, and where the *trace norm* of an operator $X$ is defined as $\|X\|_1 = \mathrm{Tr}\sqrt{X^*X}$.

### 4.2  Definition and proof of privacy

Our protocol provides the same level of security as the one-time pad, that is, it provides perfect (information-theoretic) privacy. The rest of this section formalizes the definition of privacy based on simulations and gives

3

a proof based on the technique of giving an equivalent, entanglement-based protocol. For our definition of privacy, we have used notions similar to those introduced by Watrous in the context of quantum zero-knowledge interactive proof systems [Wat09]. We use as proof technique the method of transforming a qubit-based protocol into an equivalent protocol that is more easily proved secure, but that involves entanglement. This technique is attributed to Shor and Preskill [SP00], who used it in the context of proving the security of the BB84 [BB84] quantum key exchange protocol, and has since appeared in the context of quantum message authentication [BCG+02], cryptography in the bounded-quantum-storage model [DFSS05], and in recent independent work on delegated quantum computation [DFPR13].

Formally, a protocol for delegated computation is specified by a pair $(C, S)$ representing an honest client and an honest server (without loss of generality, both parties are quantum). As the client is always honest, the security property concerns interactions between pairs $(C, S')$ where $S'$ deviates arbitrarily from $S$. At the onset of the protocol, both parties agree on the classical input $q$ which determines the general quantum circuit to be executed as an ordered series of gates acting on specified wires. The structure of the interaction between $C$ and $S$ is thus determined by $q$. At the same time, a quantum input $\rho_{\text{in}} \in \text{D}(\mathcal{C} \otimes \mathcal{S})$ is distributed, $C$ receiving the register in $\mathcal{C}$ and $S$ receiving the register in $\mathcal{S}$. A cheating server $S'$ is any quantum computational process that interacts with $C$ according to the message structure determined by $q$. By allowing $S'$ access to the input register $\mathcal{S}$, we explicitly allow $S'$ to share prior entanglement with $C$'s input; this also models any *prior* knowledge of $S'$ and formalizes the notion that the protocol cannot be used to *increase* knowledge.

Let $\mathcal{Z}$ denote the output space of $S'$ and let $\Phi_q : \text{L}(\mathcal{S}) \to \text{L}(\mathcal{Z})$ be the mapping induced by the interaction of $S'$ with $C$. Security is defined in terms of the existence of a *simulator* $\mathscr{S}_{S'}$ for a given server $S'$, which is a general quantum circuit that agrees with $S'$ on the input and output dimensions. Such a simulator does not interact with $C$ or with any other resources, but simply induces a mapping $\Psi_q : \text{L}(\mathcal{S}) \to \text{L}(\mathcal{Z})$ on each input $q$. Informally, $(C, S)$ is private if the two mappings, $\Phi_q$ and $\Psi_q$ are indistinguishable for every choice of $q$ and every choice of $\rho_{\text{in}}$. The intuitive argument for privacy is that any such simulator $\mathscr{S}_{S'}$, by definition, does not have access (even indirectly) to the client's register $\mathcal{C}$, and thus necessarily cannot extract any information from $\mathcal{C}$ (this is called the *ideal* world scenario). The mapping $\Phi_q$ models the cheating behaviour of $S'$ as a quantum channel. The security notion is that there should exist, for every cheating server $S'$ in the actual protocol (this is called the *real* world scenario), a simulator $\mathscr{S}_{S'}$ that induces an indistinguishable mapping. Since the simulator's mapping models a private protocol, the mapping of the original server $\mathscr{S}_{S'}$ must also model a private protocol. The above is the familiar ideal/real-world simulation-based security paradigm, simplified because the server does not have access to any ideal functionality, modelling the notion of privacy that the server should "learn nothing". Allowing for an $\epsilon$ amount of leakage, we formalize this as Definition 1.

**Definition 1.** *A protocol $(C, S)$ for a delegated quantum computation is $\epsilon$-private if for every server $S'$ there exists a simulator $\mathscr{S}_{S'}$ such that for every classical input $q$,*

$$\|\Phi_q - \Psi_q\|_\diamond \leq \epsilon, \tag{7}$$

*where $\Phi_q$ is the mapping induced by the interaction of $S'$ with the client $C$ on input $q$ and $\Psi_q$ is the mapping induced by $\mathscr{S}_{S'}$ on input $q$.*

Taking $\epsilon = 0$ gives the strongest possible security against a malicious server: it does not allow for even an $\epsilon$ amount of leakage, and allows the server to deviate arbitrarily (without imposing any computational bounds). Theorem 1 states that this is the level of privacy achieved in our protocol.

Our proof technique is to construct a sequence of protocols. For clarity, we refer to the protocol in the main paper as **Protocol 1**. A high-level sketch of the proof is that we modify the behaviour of the client in the main protocol (**Protocol 1**) in a way that the effect of the protocol is unchanged, yet the client delays introducing her input into the protocol until after her interaction with the server has ended (this makes the simulation almost trivial because no access to the client's input is required). In order to do so, we describe below an entanglement-based protocol (**Protocol 2**) as well as a delayed-measurement protocol (**Protocol 3**).

**Theorem 1.** *Protocol 1 is an $\epsilon$-private protocol for delegated quantum computation, with $\epsilon = 0$.*

*Proof.* Fix a value for $q$. We construct a simulator $\mathscr{S}_{S'}$ by giving instructions how to prepare messages that replace the messages that the client $C$ would send to the server $S'$ in the real protocol. Privacy follows since we will show that these transmissions are identical to those in the real protocol.

We first consider **Protocol 2**, which is an entanglement-based version of **Protocol 1**. In **Protocol 2**, we modify how the client prepares her messages, without modifying the server's actions or the effect of the protocol. Thus, the preparing and sending of an encrypted quantum register is replaced by an equivalent teleportation-based protocol, as given in Figure 1. Also, the R-gate protocol is replaced by an equivalent protocol as given in Figure 3. The protocol of Figure 3 can be seen to be correct via an intermediate protocol (Figure 2), in which the classical bit $x$ from the client to the server becomes a uniformly random bit; this transformation is possible because in the R-gate protocol, $x = a \oplus y$ with $y$ a random bit. Then choosing $x$ to be random and $y = a \oplus x$ gives an equivalent protocol. The final entanglement-based protocol of Figure 3 is seen to be correct via the circuit identity given in Figure 4. The remaining protocols for stabilizer circuit elements (Clifford gates, qubit preparation and measurements) are non-interactive and thus unchanged in **Protocol 2**.

The main advantage of considering **Protocol 2** instead of **Protocol 1** is that we can delay all the client's measurements (in Figures 1 and 3) until the output register is returned, without affecting the computation or the server's view of the protocol (because actions on different subsystems commute); call the result **Protocol 3**. In this delayed-measurement protocol, the messages from the client to the server can be chosen *before* any interaction with the server, and are thus clearly independent of the actions of $S'$.

For any cheating server $S'$ we thus construct a simulator $\mathscr{S}_{S'}$ that performs the following interactive process that emulates the real-world execution of the protocol between $S'$ and $C$: the client's operations are the same as the client's operations in **Protocol 3**, *but no measurements are performed.* (Thus, $\mathscr{S}_{S'}$ does not require access to the input register $\mathcal{C}$.) The server's operations are the same as given by $S'$. We now need to show that $\mathscr{S}_{S'}$ induces the same mapping, $\Phi_q$ as the mapping $\Psi_q$ induced by $S'$. In order to do this, note that in the construction above, $\mathscr{S}_{S'}$ actually prepares the same transmissions as would $C$ in **Protocol 1** interacting with $S'$ on any input $\rho_{\text{in}}$. It follows that running $S'$ on these transmissions (*i.e.* executing $S'$ as given by its description) will induce the same mapping as $S'$ in the real protocol, and thus $\|\Phi_q - \Psi_q\|_\diamond = 0$. □

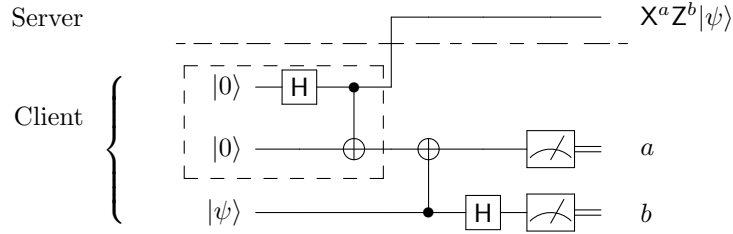

**Figure 1** – Protocol to encrypt and send a qubit using teleportation[BBC+93]. The circuit in the dashed box prepares an EPR-pair.

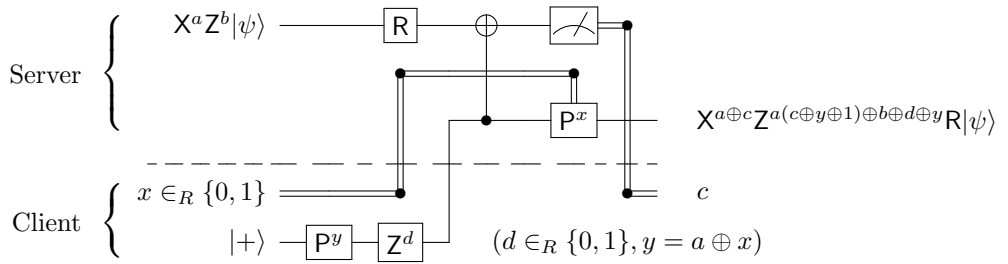

**Figure 2** – Intermediate Protocol for an R-gate. Compared to Figure 1g in the main text, the classical message from the client to the server is chosen uniformly at random. This protocol performs the same computation as the protocol in Figure 1g in the main text.

**Figure 3** – Entanglement-based protocol for an R-gate. This protocol performs the same computation as the protocols in Figure 1g in the main text and Figure 2. The circuit in the dashed box prepares an EPR-pair.

**Figure 4** – Circuit identity: entanglement-based circuit that prepares a qubit $\mathsf{Z}^d \mathsf{P}^y |+\rangle$ for uniformly random bits $y$ and $d$ (here, $y$ is chosen uniformly at random, and $d$ is determined by the measurement). The circuit in the dashed box prepares an EPR-pair.

## 5 Ideal process matrices

Below we show the ideal $\chi$ matrices for each of the gates in the universal set $\{\mathsf{X}, \mathsf{Z}, \mathsf{H}, \mathsf{P}, \mathsf{R}, \mathsf{CNOT}\}$. In the main manuscript, we present $\chi$ matrices graphically. The following figures show the ideal $\chi$ matrices corresponding to those in Figures 3 and 4 in the manuscript. Process matrices are most commonly expressed in the basis of Pauli operators, $\{\mathbb{1}, X, Y, Z\}$ for single-qubit processes, and $\{\mathbb{1} \otimes \mathbb{1}, \mathbb{1} \otimes X, ..., Z \otimes Z\}$ for two-qubit processes. Here we write process matrices as having a trace of unity, though another common convention is for the trace to equal the dimension of the system (eg., 2 for a single-qubit process). Pauli operators are straightforward to write in this basis as they only have one element in the $\chi$ matrix. Recall that the $\mathsf{X}$ Pauli operator has the following action: $\mathsf{X}|j\rangle \mapsto |\bar{j}\rangle, j \in \{0, 1\}$. Its process matrix is given by

$$\chi_{\mathsf{X}} = \begin{pmatrix} 0 & 0 & 0 & 0 \\ 0 & 1 & 0 & 0 \\ 0 & 0 & 0 & 0 \\ 0 & 0 & 0 & 0 \end{pmatrix} \tag{8}$$

The $\mathsf{Z}$ Pauli operator has the following action: $\mathsf{Z}|j\rangle \mapsto (-1)^j |j\rangle, j \in \{0, 1\}$, and has process matrix:

$$\chi_{\mathsf{Z}} = \begin{pmatrix} 0 & 0 & 0 & 0 \\ 0 & 0 & 0 & 0 \\ 0 & 0 & 0 & 0 \\ 0 & 0 & 0 & 1 \end{pmatrix} \tag{9}$$

Other gates such as the Hadamard, can be written as coherent superposition of Pauli operators, $\mathsf{H} = \frac{1}{\sqrt{2}}(X + Z)$, and will have off-diagonal terms in the $\chi$ matrix, as seen below:

$$\chi_{\mathsf{H}} = \frac{1}{2} \begin{pmatrix} 0 & 0 & 0 & 0 \\ 0 & 1 & 0 & 1 \\ 0 & 0 & 0 & 0 \\ 0 & 1 & 0 & 1 \end{pmatrix} \tag{10}$$

The phase, P, gate has the following action: $P|j\rangle = i^j|j\rangle, j \in \{0,1\}$. It can also be written as a superposition of the identity, $\mathbb{1}$, and Z Pauli operators: $P = \frac{1}{\sqrt{2}}(\mathbb{1} - iZ)$. Its process matrix is given by

$$\chi_P = \frac{1}{2} \begin{pmatrix} 1 & 0 & 0 & i \\ 0 & 0 & 0 & 0 \\ 0 & 0 & 0 & 0 \\ -i & 0 & 0 & 1 \end{pmatrix} \tag{11}$$

Recall, the R gate has the following action: $R|j\rangle = \left(e^{i\frac{\pi}{4}}\right)^j|j\rangle, j \in \{0,1\}$. Again, it can be written as a superposition Pauli operators: $R = (\cos\frac{\pi}{8}\mathbb{1} - i\sin\frac{\pi}{8}Z)$. Its process matrix is given by

$$\chi_R = \begin{pmatrix} \frac{1}{4}(2+\sqrt{2}) & 0 & 0 & \frac{i}{2\sqrt{2}} \\ 0 & 0 & 0 & 0 \\ 0 & 0 & 0 & 0 \\ -\frac{i}{2\sqrt{2}} & 0 & 0 & \frac{1}{4}(2-\sqrt{2}) \end{pmatrix} \tag{12}$$

The last gate we look at is the CNOT, which has the following action $\text{CNOT}|j\rangle|k\rangle \mapsto |j\rangle|k \oplus j\rangle, j,k \in \{0,1\}$. In the basis of Pauli operators, it is written as $\text{CNOT} = \frac{1}{2}(\mathbb{1}\otimes\mathbb{1} + \mathbb{1}\otimes X + Z\otimes\mathbb{1} + Z\otimes X)$. Its process matrix is

$$\chi_{\text{CNOT}} = \frac{1}{4}\begin{pmatrix} 1 & 1 & 0 & 0 & 0 & 0 & 0 & 0 & 0 & 0 & 0 & 0 & 1 & -1 & 0 & 0 \\ 1 & 1 & 0 & 0 & 0 & 0 & 0 & 0 & 0 & 0 & 0 & 0 & 1 & -1 & 0 & 0 \\ 0 & 0 & 0 & 0 & 0 & 0 & 0 & 0 & 0 & 0 & 0 & 0 & 0 & 0 & 0 & 0 \\ 0 & 0 & 0 & 0 & 0 & 0 & 0 & 0 & 0 & 0 & 0 & 0 & 0 & 0 & 0 & 0 \\ 0 & 0 & 0 & 0 & 0 & 0 & 0 & 0 & 0 & 0 & 0 & 0 & 0 & 0 & 0 & 0 \\ 0 & 0 & 0 & 0 & 0 & 0 & 0 & 0 & 0 & 0 & 0 & 0 & 0 & 0 & 0 & 0 \\ 0 & 0 & 0 & 0 & 0 & 0 & 0 & 0 & 0 & 0 & 0 & 0 & 0 & 0 & 0 & 0 \\ 0 & 0 & 0 & 0 & 0 & 0 & 0 & 0 & 0 & 0 & 0 & 0 & 0 & 0 & 0 & 0 \\ 0 & 0 & 0 & 0 & 0 & 0 & 0 & 0 & 0 & 0 & 0 & 0 & 0 & 0 & 0 & 0 \\ 0 & 0 & 0 & 0 & 0 & 0 & 0 & 0 & 0 & 0 & 0 & 0 & 0 & 0 & 0 & 0 \\ 0 & 0 & 0 & 0 & 0 & 0 & 0 & 0 & 0 & 0 & 0 & 0 & 0 & 0 & 0 & 0 \\ 0 & 0 & 0 & 0 & 0 & 0 & 0 & 0 & 0 & 0 & 0 & 0 & 0 & 0 & 0 & 0 \\ 1 & 1 & 0 & 0 & 0 & 0 & 0 & 0 & 0 & 0 & 0 & 0 & 1 & -1 & 0 & 0 \\ -1 & -1 & 0 & 0 & 0 & 0 & 0 & 0 & 0 & 0 & 0 & 0 & -1 & 1 & 0 & 0 \\ 0 & 0 & 0 & 0 & 0 & 0 & 0 & 0 & 0 & 0 & 0 & 0 & 0 & 0 & 0 & 0 \\ 0 & 0 & 0 & 0 & 0 & 0 & 0 & 0 & 0 & 0 & 0 & 0 & 0 & 0 & 0 & 0 \end{pmatrix} \tag{13}$$

We compare the server's results from quantum process tomography to the compleltely depolarizing channel. The the action of this channel on a density matrix $\rho$ is $\rho \mapsto \frac{1}{4}(\mathbb{1}\rho\mathbb{1} + X\rho X + Y\rho Y + Z\rho Z)$. Its process matrix is then diagonal,

$$\chi_{\text{dep.}} = \frac{1}{4}\begin{pmatrix} 1 & 0 & 0 & 0 \\ 0 & 1 & 0 & 0 \\ 0 & 0 & 1 & 0 \\ 0 & 0 & 0 & 1 \end{pmatrix} \tag{14}$$

Similarly for two-qubit processes, the completely depolarizing channel has only diagonal elements, and so $\chi_{\text{dep.}} = \frac{1}{16}\mathbb{1}_{16}$, where $\mathbb{1}_d$ is the $d$-dimensional identity matrix.

## 6 Analysis of experimental security

Our experimental implementation serves as a proof-of-principle for our protocol of quantum computing on encrypted data, and as such is bound by some limitations. In this section we discuss methods of the server might use to break the client's encryption via experimental imperfections, or multiple photon emissions from the source.

## 6.1 Imperfect encryption

As is the case for the quantum server, our experimental client suffers from imperfect gate fidelities. In particular, the encryption operation differs from the ideal encryption $\mathsf{X}^a\mathsf{Z}^b$. Performing a quantum process tomography using the same method as in the main text, we found that the client's single-qubit encryption operations had the following fidelities, $F_{ab}$:

| $a$ | $b$ | $F_{ab}$ |
|---|---|---|
| 0 | 0 | 0.976 |
| 0 | 1 | 0.986 |
| 1 | 0 | 0.989 |
| 1 | 1 | 0.957 |

The ideal encryption operation on an input state $\rho$, when mixed over the different cases of $a$ and $b$, can be expressed by

$$\rho_{\text{enc}}^{\text{id}} = \frac{1}{4}\sum_{ab} \mathsf{X}^a\mathsf{Z}^b\rho\mathsf{Z}^b\mathsf{X}^a. \tag{15}$$

This can be written as a diagonal process matrix $\chi_{\text{enc}}^{\text{id}} = \frac{1}{4}\mathbb{1}$. Assuming that any departure from the ideal encryption operation leaves the original state $\rho$ as is, we express it by a similar mapping

$$\rho_{\text{enc}}^{\text{exp}} = \frac{1}{4}\sum_{ab}\left(F_{ab}\cdot\mathsf{X}^a\mathsf{Z}^b\rho\mathsf{Z}^b\mathsf{X}^a + (1-F_{ab})\cdot\rho\right). \tag{16}$$

This can be written as a diagonal process matrix $\chi_{\text{enc}}^{\text{exp}}$ with entries $\{\frac{1}{4}(1-F_{01}-F_{10}-F_{11}), \frac{1}{4}F_{01}, \frac{1}{4}F_{10}, \frac{1}{4}F_{11}\}$.

In the theoretical discussion of Section 4, we use the diamond norm as a distance measure. However, since the calculating the diamond norm from experimental data is computationally intensive, involving a maximization over an arbitrary amount of ancilla systems, we instead compare the experimental encryption operation against the ideal using the maximal trace distance measure. This is defined as $\mathcal{D} = \max_\rho \frac{1}{2}\text{Tr}\left|\rho_{\text{enc}}^{\text{exp}} - \rho_{\text{enc}}^{\text{id}}\right|$, where $|A| = \sqrt{A^\dagger A}$ and the input state $\rho$ is in a two-dimensional Hilbert space. Operationally, $\mathcal{D}$ is the highest probability of distinguishing between the experimental and ideal encryption operations when inputting the appropriate single qubit state. We calculate our experimental encryption operation to have a maximum trace distance of 0.014 away from the ideal, where the maximizing input state is close to $\rho = |R\rangle\langle R|$.

## 6.2 Multi-photon emissions

Our implementation of the full protocol demonstrates that, with only modest additional resources, the required gates can be performed. While the protocol is secure, experimental imperfections can provide the server with side channels from which they can learn information about the client's key. In our experiment the primary security flaw is caused by emission of multiple photon pairs.

The Clifford gates we study are secure against the server obtaining multiple copies of the encrypted qubits. Even if the server obtains multiple copies of the encrypted data the server cannot learn the key. The R gate, however, presents a malicious server with an opportunity to discover a portion of the client's key. To carry out the R gate, the client must send the server an auxiliary qubit that is used to hide the value of the encryption bit $a$ from the server. The auxiliary qubit is prepared in one of the four BB84 states, $\mathsf{Z}^d\mathsf{P}^y|+\rangle$, according to the randomly chosen classical bits $y$ and $d$. If the server obtains multiple copies of this auxiliary qubit, then the value of $a$ can be determined and the key discovered.

To implement the fast, random, switching between BB84 states during the experiment we used Pockels' cells which performed the unitaries $\mathsf{P}^y$ and $\mathsf{Z}^d$. Although we reduce the photon rates emitted from the source, and switch the Pockels' cells quickly (1 MHz), we still have the possibility that multiple photons can be present during a single setting of the Pockels' cells. Based on single and coincident photon detections in each arm of the experiment, we can estimate the average number of photons passing through the Pockels' cells during a single setting. After losses in fibre-coupling, the CNOT gate, and in the detectors we measure the efficiency of the auxiliary photon path and estimate the number of photons present in the Pockels' cell to be on average 0.96 during a single setting of $y$ and $d$. This confirms that the Pockels' cell switching, paired with reduced photon rates, is approaching what is required for security against the server. Faster switching

of the Pockels' cells, or equivalently, employing a laser with a repetition rate matching that of the Pockels' cells (such as a pulse picker), along with better photon sources, and a higher system efficiency would allow this security to be improved significantly.

Unless the optical setup has a high efficiency, the client could be leaking information to the server. A primary stumbling block to the efficiency of the optical setup is the implementation of the CNOT gate. The configuration has a success probability of 1/9 and the server obtains many additional photons without the client's knowledge, which can in principle be used to break the encryption. While this presents a major security issue, we emphasize that the issue is strictly technological and not fundamental. A photonic implementation is useful for a proof-of-principle of our protocol, but a high-fidelity, deterministic implementation of the CNOT gate will drastically improve the experimental security.

This is the first demonstration of a delegated quantum computation that has security. Previous work [BKB$^+$12] did not employ fast switching and could be broken by a server who has access to more than one copy of the qubit and could be broken due to the server's access to several identical copies of the auxiliary qubits. A security proof which takes into account the amount of information leaked to the server, e.g., establishing entropic bounds, is beyond the scope of the current work and will be the subject of future research.



\end{document}